# From Fragmentation to Liberation

*Nick Merrill*

*October 9, 2021*

The Internet has never floated freely, untethered from political realities. Nor has it ever been truly global. Nevertheless, a growing panic has emerged around "Internet fragmentation," evoking the threat of a so-called "splinternet." And for good reason. A fragmented Internet carries consequences for free speech, international business, trade, and more. Insofar as the Internet mediates access to goods and services, variations across territorial space will radically shape life chances, [1] particularly those of the 3.6 billion people who are just now getting online. [2] It's no wonder that the topic has spooked businesses, policymakers, activists, and militaries alike. It hooks into larger debates about "decoupling," trade protectionism, a rising China, and concerns about liberalism under threat globally. Yet we don't have good theories to explain why internet fragmentation is happening in the specific ways we observe it to be occurring. [3] Consider the following cases:

- The EU's General Data Protection Framework (GDPR), a legal framework for handling data that renders the Internet a bit different in the EU than in the U.S.

- The "Great Firewall of China," which censors most of the western Internet (for example, Google).

- Russia's data sovereignty laws, which mandate data about Russians be housed in Russian servers, culminating in a drill that disconnected Russia from the global Internet.

Why did these different actors employ these particular tactics? Why is Russia's strategy so different from China's? Without answers to questions like these, it's difficult to imagine how decision-makers worldwide could ever react to a changing Internet and (entangled with it) a changing geopolitical landscape. [4]

In this paper, I argue that "Internet fragmentation" as a phenomenon is only meaningful in the context of the U.S.'s hegemonic control over the Internet. I propose a broader and, I argue, more richly predictive frame: Internet conflict. I show how this frame provides fresh analytical purchase to some of the questions I list above, using it to contextualize several apparently distinct phenomena. I conclude by arguing that only one question gives this analytical frame, or any other, a higher purpose: *what particular interventions to Internet governance can produce meaningfully liberatory outcomes?* Any descriptive framework is only useful insofar as it can be mobilized to answer this normative question.

---

[1] Marion Fourcade and Kieran Healy. Classification situations: Life-chances in the neoliberal era. *Accounting, Organizations and Society*, 38(8):559–572, 2013

[2] Anne Jonas and Jenna Burrell. Friction, snake oil, and weird countries: Cybersecurity systems could deepen global inequality through regional blocking. *Big Data Society*, 6(1):2053951719835238, 2019

[3] Our early quantitative research revealed diverse patterns of fragmentation around the world, complicating reductive notions of "free" and "closed" Internets.

[4] The Internet constitutive of—that is, both shapes and reflects—geopolitics.

Frédérick Douzet. La géopolitique pour comprendre le cyberespace. *Hérodote*, (1):3–21, 2014

*Background*

Discussions of Internet fragmentation (roughly, the idea that the Internet is becoming increasingly different in different places) elude a basic truth: that the United States hegemonically controls the Internet, both directly through physical control over (here meaning the potential use of force on) core Internet infrastructure, and indirectly via its jurisdiction over the tech companies that provision it. In fact, using court orders, the U.S. could block access to almost all content on the Internet. Thanks to the wide range of core Internet infrastructure provisioned by U.S.-domiciled actors, the particular blocking strategies available to the U.S. government could make affected content inaccessible globally, not just in the United States. I draw these findings from quantitative data on the market share of various tech companies who provision core Internet infrastructure. [5] Our findings, while complex and multifaceted, can be illustrated succinctly through two cases: domain names and certificate authorities.

[5] I collected these data in partnership with the Internet Society and W3Techs. You can find our methodology and the code we used to collect our data on [GitHub](). Please contact me if you would like to access to the full dataset and various visualization tools.

*Domain names*

Domain names are human readable names (e.g., "nytimes.com") that map to machine-readable IP addresses in the global Domain Name System (DNS). Domain names allow us to access content using memorable addresses. Domain names come in the form:

`members.parliament.uk`

We interpret this name hierarchically: each name is "contained within" the name that follows. So, in this example ".uk" is the highest zone in the hierarchy: it is the top-level domain or TLD. Domain registry backends have the highest level of authority over what domain names (e.g., "nytimes.com") map to which IP addresses within its top-level domain (i.e., ".com").

Since there is only one registry backend per top-level domain, a state with jurisdiction over a particular registry backend can use its legal power (or, theoretically, brute force) to "seize" any domain name. Effectively, this amounts to redirecting requests from a given domain to content of the state's choice.

The United States federal government already blocks access to content using domain seizures. So far, the United States has taken down over 1.2 million websites using this technique through an ICE-led operation called Operation In Our Sites, most of the domains seized without due process. [6] And the U.S.'s use of domain seizures is expanding. Recently, the U.S. DOJ [seized the domain of an Iranian news sites]() after charging them with spreading "disinformation."

[6] Karen Kopel. Operation seizing our sites: How the federal government is taking domain names without prior notice. *Berkeley Tech. LJ*, 28:859, 2013

Now, consider this: the world's most popular top-level domain names (e.g., ".com" and ".net" are provisioned by domain registry backends based in the United States. As of August 19, 2021, 65.9% of Alexa-ranked websites are accessible only via top-level domains whose registry backends are in U.S. jurisdiction. The U.S. Department of Justice (DOJ) could use domain seizures to block access to approximately 64.9% of the content on the web—that is, to make that content inaccessible to everyone.

*Certificate authorities*

If you've ever noticed the lock icon in the address bar of your web browser, then you're already familiar with Transport Layer Security (TLS). Through a system of digital "certificates" granted by Certificate Authorities (CAs), Internet traffic can be encrypted during transit. This prevents both eavesdropping and so-called person-in-the-middle attacks, in which content is silently altered during transit.

To illustrate a person-in-the-middle attack: imagine entering "nytimes.com" and hitting enter. The page loads. There's a lock in the corner of your browser. It looks just like the New York Times. But, unbeknownst to you, it's not the real New York Times: it's a carefully-crafted replica made by Chinese intelligence. A few stories have been removed. Perhaps a few have been added. And you'd have no way of knowing.

A sufficiently well-resourced actor could do just this. By altering the DNS record (as described in Section ) and subsequently issuing a valid certificate for that website via a certificate authority in their jurisdiction, [7] an attacker could serve a forged website, and the lock would still appear in your browser! [8] If we can't trust media or the cryptographic mechanisms that underpin the delivery of content on the web, we're left with a "truth doomsday" scenario, the potential impacts of which are hard to see the bottom of.

Now consider, as of August 19 2021, 97.3% of Alexa-ranked websites use CAs based in the United States. The U.S. has not yet exercised its jurisdiction over CAs as it has over domain registrars and registry backends. But it need not use its power to block content outright. Through court orders, the United States DOJ could perform what I call "soft blocking:" by demanding CAs to revoke certificates, or asking web browser vendors (like Google) to revoke trust in a CA, users will see a scary notice when they attempt to visit a targeted website (Figure 1).

---

[7] Note: A CA can issue a valid certificate for any domain, even one for which it did not previously issue a certificate.

[8] David D. Clark. Control point analysis. *SSRN Electronic Journal*, 2012. DOI: 10.2139/ssrn.2032124. URL https://doi.org/10.2139/ssrn.2032124

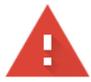

Figure 1: The error you'll see if you try to visit a site with a revoked certificate.

*The state of the Internet is one of hegemonic control*

This discussion only scratches the surface of a deep and complex topic. I have not mentioned, for example, content delivery networks (CDNs), a highly-concentrated piece of internet infrastructure that can (and does) allow fine-grained blocking of web content. My goal is not to suggest that the U.S. will opt to perform any of the tactics I have described. Rather, my goal is to demonstrate what the U.S. federal government apparatus could do, if it chose to. The critical point is this: The potential power of the U.S. over the Internet is astounding, more complete than most people appreciate. Broadly speaking, the content available on the Internet is available because the U.S. government allows it to be.

*From fragmentation to conflict*

Where received wisdom frames the "splinternet" as rogue actors sabotaging a common good, I propose an alternative and, I argue, richer frame: that various actors—Westphalian states, tech companies, and popular movements (e.g., cryptocurrency projects)—opportunistically contest U.S. hegemony to achieve particular strategic goals. What we call "internet fragmentation" is the observable effect of that contestation.

This frame is both richer and more predictive than narratives that center "the Splinternet" per se. For example, this frame helps to explain Russia's data sovereignty laws, mentioned above. Why is Russia disconnecting itself from the global Internet? What is its intel-ligence apparatus hoping to achieve? The frame of Internet conflict makes clear one possibility: Russian intelligence may want to have the option to attack core internet infrastructure. Such an attack would bring down the Internet for everyone—except for them, thanks to their ability to disconnect from it.

This frame also helps to "collapse" a few debates that appear distinct. For example, consider the so-called "techlash," in which states (particularly the U.S. are increasingly attempting to rein in tech companies, in hopes of aligning [9] their policies with states' goals. I argue: "Internet fragmentation" is to the conflict among states as this "techlash" is to the conflict between states and tech companies. As "walled gardens" are to conflict among tech companies. All describe power struggles "in" the Internet, as a domain of conflict. And all indicate a U.S. whose hegemonic power faces active challenges by state and non-state actors alike.

[9] Milton Mueller. *Will the internet fragment?: Sovereignty, globalization and cyberspace*. John Wiley & Sons, 2017

This frame not only helps us understand the present; it also helps us imagine possible futures. Debates around "Internet fragmentation" envision a web of "splintered" internets, each in self-contained national and/or corporate siloes. That vision eludes a highly likely, though infrequently discussed, scenario: an Internet that's global, but globally censored. [10]

I am not the first to pick at the concept of Internet fragmentation. In his 2017 monograph, Milton Mueller proposes replacing the term "Internet fragmentation" with the notion that Westphalian states are trying to "re-align" control of communication systems with their jurisdictional boundaries. [11] But there is no "re" in this "re-"alignment. As long as the Internet has existed, it has been "aligned" with the goals of the United States and corporations who jointly enacted U.S. interests throughout the 20th century. [12] Only recently, most visibly in the wake of the 2016 presidential election, have their interests substantively diverged.

[10] Imagine, for example, an "American Great Firewall" (a "Great Border Wall?") that censors content globally, using a mix of domain seizures, DNS cache poisoning, court orders on CDNs, etc. This scenario depicts an "offensive realist" reading of the Internet conflict frame, in which every actor seeks total domination. Arguably, it is the U.S.'s laissez-faire approach that has enabled its dominance. That calculus may someday change.

[11] Milton Mueller. *Will the internet fragment?: Sovereignty, globalization and cyberspace*. John Wiley & Sons, 2017

[12] Tung-Hui Hu. Truckstops on the Information Superhighway: Ant Farm, SRI, and the Cloud. *Journal of the New Media Caucus*, 2014. URL http://median.newmediacaucus.org/art-infrastructures-h...

Nor or states the only actors seeking to align communication systems. Technology companies in the United States have long sought, and in many ways succeeded, in aligning the Internet with their strategic goals. This ad-hoc conflict and cooperation, an ongoing process of alignment among various actors, has been a constant throughout the Internet's history, and will remain one for its foreseeable future.

*From conflict to liberation*

Currently, the U.S.'s global power projection, both military and cultural, is faltering. On the Internet, this manifests as more frequent and more visible misalignment between state interests and domestic tech companies.

The question then becomes: what Internet comes next? A global Internet led by China? By a single tech company? Or by a consortium of states and tech companies, working in cooperation, as they already to do in the case of companies like Palanatir, to "securitize"?[13] And what will the transition to the next Internet look like? Will the ensuing conflict be multilateral and purposeful, or power-based and ad-hoc? Will it ever arrive at a stable configuration, or will it devolve into eternal conflict on a barren and unusable web?

Of all of those admittedly very interesting questions, the question I'm left with is this: *what particular interventions to Internet governance can produce meaningfully liberatory outcomes?* I think this is the only meaningful question in all of Internet governance. Any descriptive framework we come up with is only useful insofar as we can mobilize toward this normative question.

Yet "liberation" and "Internet governance" seem no longer to mix, at least not in the circles in which I run.[14] I understand why. Today, the techno-utopian promises of the 1990s have completed their rot: the Internet represses globally. Censorship, shutdowns, surveillance; not to mention gig work, pre-automation, and other regimes of labor subjugation discussions of Internet freedom so strategically omit. This Internet has delivered new vectors of power to the usual suspects: state and capital interests.[15]

Today, thoughtful Internet types are more interested in defending ourselves against the worst abuses of ruling interests (for example, protecting against surveillance) than in trying to subvert those interests. We dare not imagine the potential of technically-instantiated networks to deliver, alongside and through ongoing social practice, new ways of being and living; challenges to dominant logics of capital accumulation; challenges to colonial power.

Who does this reactive approach ultimately benefit? Are we really so pessimistic about the ability of networked technologies to assist in liberatory struggles? It seems to me that a communications network that is *ideologically* trustworthy a necessary pre-requisite to any liberatory struggle we might hope to achieve in our lifetimes. From paranets for protesters to streetnets in Havana, some of the "fragmented" Internets we view today are testaments to that fact. [16] That network will be an internet. Just maybe not this internet.

[13] Perhaps the U.S. government's primary complaint against Facebook is that it fails to participate satisfactorily in this regime.

[14] The people who do talk about liberation are more likely to be talking about bitcoin which, if it does stand for liberation, stands for a very different type from the kind I'm interested in.

[15] I have little doubt that blockchain developers will deliver their technologies to the same masters, if that wasn't their intention all along.

[16] Abigail Z Jacobs and Michaelanne Dye. Internet-human infrastructures: Lessons from havana's streetnet. *arXiv preprint arXiv:2004.12207*, 2020